# Virtual Screening of Plant Metabolites against Main protease, RNA-dependent RNA polymerase and Spike protein of SARS-CoV-2: Therapeutics option of COVID-19

**Md Sorwer Alam Parvez[1&] , Kazi Faizul Azim[2,3&], Abdus Shukur Imran[2,4&], Topu Raihan[1], Aklima Begum[2], Tasfia Saiyara Shammi[2], Sabbir Howlader[5], Farhana Rumzum Bhuiyan[6,7], Mahmudul Hasan[2,4]**

[1]Department of Genetic Engineering and Biotechnology, Shahjalal University of Science and Technology, Sylhet, Bangladesh

[2]Faculty of Biotechnology and Genetic Engineering, Sylhet Agricultural University, Sylhet, Bangladesh

[3]Department of Microbial Biotechnology, Sylhet Agricultural University, Sylhet, Bangladesh

[4]Department of Pharmaceuticals and Industrial Biotechnology, Sylhet Agricultural University, Sylhet, Bangladesh

[5]Department of Applied Chemistry and Chemical Engineering, University of Chittagong, Chittagong, Bangladesh

[6]Department of Botany, University of Chittagong, Chittagong, Bangladesh

[7]Laboratory of Biotechnology and Molecular biology, Department of Botany, University of Chittagong, Chittagong, Bangladesh

[&] ***authors contributed equally to this work***

***Corresponding Author**

Mahmudul Hasan
Assistant Professor
Department of Pharmaceuticals and Industrial Biotechnology
Faculty of Biotechnology and Genetic Engineering
Sylhet Agricultural University, Sylhet, Bangladesh
Email: mhasan.pib@sau.ac.bd
Mobile: 008801723698461

# Virtual Screening of Plant Metabolites against Main protease, RNA-dependent RNA polymerase and Spike protein of SARS-CoV-2: Therapeutics option of COVID-19

## Abstract

Covid-19, a serious respiratory complications caused by SARS-CoV-2 has become one of the global threat to human healthcare system. The present study evaluated the possibility of plant originated approved 117 therapeutics against the main protease protein (MPP), RNA-dependent RNA polymerase (RdRp) and spike protein (S) of SARS-CoV-2 including drug surface analysis by using molecular docking through drug repurposing approaches. The molecular interaction study revealed that Rifampin (-16.3 kcal/mol) were topmost inhibitor of MPP where Azobechalcone were found most potent plant therapeutics for blocking the RdRp (-15.9 kcal /mol) and S (-14.4 kcal/mol) protein of SARS-CoV-2. After the comparative analysis of all docking results, Azobechalcone, Rifampin, Isolophirachalcone, Tetrandrine and Fangchinoline were exhibited as the most potential inhibitory plant compounds for targeting the key proteins of SARS-CoV-2. However, amino acid positions; H41, C145, and M165 of MPP played crucial roles for the drug surface interaction where F368, L371, L372, A375, W509, L514, Y515 were pivotal for RdRP. In addition, the drug interaction surface of S proteins also showed similar patterns with all of its maximum inhibitors. ADME analysis also strengthened the possibility of screened plant therapeutics as the potent drug candidates against SARS-C with the highest drug friendliness.

## 1. Introduction

COVID-19 pandemic situation has tremendously turned the entire world into a place of horrible death tragedy. SARS-CoV-2, initially named as 2019 novel coronavirus (2019-nCoV) by the World Health Organization (WHO) is the causative agent of the recent serious respiratory complications resulting the COVID-19 pandemic [1,2,3,4]. Though the symptoms of COVID-19 infection appear after an incubation period of approximately 5.2 days, the period from the onset of COVID-19 symptoms to death ranged from 6 to 41 days with a median of 14 days [5,6]. It has already completed its world tour, and around 213 countries are now experiencing the deadly scene occurred by COVID-19 including 41, 52,670 infected patients and 2, 84,536 global death cases till 12th May, 2020 [7]. The scientific community is racing to explore the effective remedy against this severe health complications, but till to date there are no any potential therapeutics have been approved for clinical use [8].

There have been few key proteins of SARS-CoV-2 that could be targeted as the vaccine or drug surface [9]. Similar to SARS and MERS, non-structural proteins (e.g. 3-chymotrypsin-like protease coronavirus main protease, papain-like protease, helicase, and RNA-dependent RNA polymerase), structural proteins (e.g. spike glycoprotein) and accessory proteins were investigated in the genome of SARS-CoV-2 where non-structural proteins constitute two-thirds of the entire genome [10]. Among the structural proteins, Nucleocapsid (N) protein is prerequisite for RNA genome assembly where Membrane (M) and Envelope (E) proteins are associated in viral assembly in the host environment [11]. Moreover, Spike (S) protein is mainly responsible for the viral entry into the host cell, and that is why spike protein is now being considered as a major therapeutic target for drug and vaccines against SARS-CoV-2 [12]. Again, The S protein interaction with the human ACE2 interface has been revealed at the atomic level, and the efficiency of ACE2 usage was found to be a main factor of coronavirus transmissibility in human to human [13]. On the contrary, coronavirus main protease (M pro) or 3C-like proteinase (3CLP) was reported for their ability to cleave the polyproteins into individual polypeptides that are required for replication and transcription [14]. The 3CLP is auto-cleaved initially from the polyproteins to become a mature enzyme leading the translation of the messenger RNA [15]. Then the 3CLP cleaves all the 11 downstream non-structural proteins. As 3CLP plays a vital role in the replication cycle of virus in the host, it has been reported as the attractive target against the human SARS virus [16]. RNA-dependent RNA polymerase, other key target protein of SARS-CoV-2 catalyses the synthesis of viral RNA possibly with the support of other non-structural proteins as co-factors [17,18,19].

The computational drug repurposing method could allow the immediate search of potential antiviral therapy in case of re-emergence of viral infections as like as COVID-19 pandemic situation [20,21,22]. Computational drug repurposing has already been used to identify promising drug candidates for other virus associated diseases like Dengue, Ebola, ZIKA, and influenza infections [23,24]. Most importantly, the SARS-CoV-2 has shown evolutionary convergent relations with SARS-CoV and MERS-CoV, and the drug repurposing methods were also applied to SARS-CoV and MERS-CoV [25,26,27]. Hence, extensive in silico studies were performed to identify potential drug candidates, for example, Prulifloxacin, Bictegravir, Nelfinavir, and Tegobuvi, were identified as repurposing candidates against COVID-19 by looking for drugs with high binding capacity with SARS-CoV main protease [28]. Again, Nelfinavir, an HIV-1 protease inhibitor was also predicted to be a potential inhibitor of COVID-19 main protease by another computational-based study [29].

However, secondary metabolites from plant origin are found to show effective defence mechanism against different deadly pathogens, and they have been widely used for conventional remedy to treat a wide range of human diseases since the ancient period of human civilization [30,31,32]. In this pandemic situation, researchers are trying find out the effective solution against COVID-19 where plant metabolites could be a promising wings for screening out potential drug candidates. Even few plant secondary metabolites have already been reported as effective against other coronaviruses [33,34]. In the present study, a total of 117 plant based drug compound were screened out to check their potentiality for blocking the three important key proteins of SARS-CoV-2. The main protease proteins, RNA-dependent RNA polymerase and spike protein of SARS-CoV-2 were employed to molecular docking study with the repurposed drug candidates from plant origin for find out the better drug option towards the COVID-19 pandemic.

## 2. Materials and Methods

### 2.1 Retrieval of SARS-CoV-2 main protease proteins, RNA-dependent RNA polymerase, spike protein and acquisition of potential natural therapeutics

PDB structures of SARS-CoV-2 main protease proteins (6LU7, 6Y2E), RNA-dependent RNA polymerase (6M71) and spike protein (6VYB) were retrieved from RCSB Protein Data Bank [35]. Moreover, a total 117 plant based drugs were collected from PubChem database (Supplementary Table 1). Alpha-ketoamide (CID 6482451) were also retrieved from the PubChem database database (https://pubchem.ncbi.nlm.nih.gov/) of NCBI [36].

### 2.2. Screening of natural therapeutics against the key viral proteins

Molecular docking is an effective approach for screening out potential therapeutics against specific drug-targets of deadly pathogens [37,38]. The crystal structure of retrieved SARS-CoV-2 proteins (complexed with inhibitors) were refined by PyMOL v2.0 software [39]. Unwanted molecules i.e. water, ions, inhibitors were removed from the viral retrieved viral protein, and further employed to molecular docking experiment with 117 natural therapeutics. AutoDock Vina software [40] to analyse the binding affinity and interactive amino acids. Alpha-ketoamide, an inhibitor SARS-CoV-2 main protease protein were used as a positive control in this study [41] and also docked against the target proteins of SARS-CoV-2. The default parameters for grid box were set to 62 A° x 71 A° x 60 A° (x, y and z) and center -25.389 A° x 15.333 A° x 56.865 A° (x, y and z) to perform the action. Moreover, 2D ligand-protein interaction diagrams were generated by LigPlot+ find out the involved amino acids with their interactive position were identified in the docked complexes [42]. The ligand molecules' interactions with the viral proteins were visualize and analyzed by Discovery Studio [43] and PyMOL v2.0 software [44].

### 2.3. Structural insights of drug surface hotspot in the viral proteins

The drug surface hotspot of SARS-CoV-2 proteins were identified by analysing the docked structures of each protein with the top most natural therapeutics by LigPlot+, PyMOL v.2.0 and Discovery Studio software. Binding patterns of Azobechalcone, Rifampin, Isolophirachalcone, Tetrandrine, Fangchinoline with SARS-CoV-2 proteins and the results were allowed for the comparative structural analysis of screened natural therapeutics. Furthermore, interactions of Alpha-ketoamide with the studied proteins were also investigated.

### 2.4. Analysis of drug likeness property of top drug candidates

The ADME (Absorption, Distribution, Metabolism and Excretion) properties of top drug candidates were analysed by SwissADME server [45]. The pharmacokinetics, drug-likeness property and medicinal chemistry were assessed [46]. Default parameters were used to evaluate various physiochemical parameters (Molar Refractivity, Molecular weight, TPSA), lipophilicity (Log Po/w (WLOGP), Log Po/w (MLOGP), Log Po/w (XLOGP3), Log Po/w, (SILICOS-IT), Log Po/w (iLOGP), Consensus Log Po/w), pharmacokinetics parameters (Log Kp; skin permeation) and water solubility of the probable drug candidates [47]. The inhibition effects of these natural therapeutics with different cytochromes P450s (CYP2C9, CYP2C19 CYP1A2, CYP3A4, CYP2D6) were also studied. In addition, admetSAR and OSIRIS Property Explorer were employed to evaluate the toxic or undesired effects (i.e. mutagenicity, tumerogenecity) of the compounds [48,49,50].

## 3. Results

### 3.1. Screening of natural therapeutics against the key viral proteins

All of the retrieved natural therapeutics were employed for molecular docking against MPP, RdRp and Spike protein of SARS-CoV-2 (Supplementary Table 2). The scoring function of AutoDock Vina was utilized to predict the interaction between the ligands (therapeutics) and the proteins. The top five inhibitors for each protein was identified based on their free binding energy. Results showed that Rifampin had the highest negative binding energy (-16.3 kcal/mol) among top MPP inhibitors (Table 1). Azobechalcone (-14.6 kcal/mol), Isolophirachalcone (-13 kcal/mol), Amentoflavone (-12.8 kcal/mol) and Cepharanthine (-12.7 kcal/mol) docked with the MPP of SARS-CoV-2 were also exhibited topmost place with a higher negative binding energy (<-12.7 kcal/mol) as well (Table 1). While interacting with RdRp of SARS-CoV-2, the most negative binding energy was scored by Azobechalcone (-15.9 kcal /mol) following by Rifampin (-15.6 kcal/mol), Tetrandrine (-13.9 kcal/mol), Biflavone (-13.7 kcal/mol), Biflavone (-13.7 kcal/mol) (Table 2). Moreover, Azobechalcone, Rifampin, Isolophirachalcone, Fangchinoline and Tetrandrine were found to be top most natural inhibitors for the spike protein of SARS-CoV-2. Azobechalcone required lowest energy (-14.4 kcal/mol) to interact with the spike protein, while Rifampin, Isolophirachalcone, Fangchinoline and Tetrandrine scored -13.7 kcal/mol , -12.8 kcal/mol, -12.6 kcal/mol and -12.5 kcal/mol respectively (Table 3). However, Azobechalcone, Rifampin, Isolophirachalcone, Tetrandrine and Fangchinoline were found to be most effective inhibitory natural compounds when the docking results were compared for all three SARS-CoV-2 proteins (Figure 2 and Table 4). All of these plant based natural inhibitors required minimum energy (not more than -12.3 kcal/mol) to interact with the studied protein molecules.

**3.2.Structural insights of drug surface hotspot in the viral proteins**

The docking pattern and interacting amino acid residues with their respective position were analyzed to unravel the binding sites of studied SARS-CoV-2 proteins. Rifampin were involved with the amino acid H41, N142, S144, C145, H163, M165, E166, D187, R188, Q189 of MPP of SARS-CoV-2 (Figure 2). The position of H41, C145, and M165 were also crucial for the binding of Amentoflavone (H41, C145, H164, M165, E166, D187, R188, Q189) and Cepharanthine (H41, N142, C145, M165, E166, D187, R188, Q189) (Table 1). Azobechalcone, the top scorer among RdRp inhibitors, were engaged by Y32, K47, L49, Y129, H133, N138, D140, T141, S709, T710, D711, K714, K780, N781 in the docked complex. Moreover, results revealed that 6 amino acid positions i.e. F368, L371, L372, A375, W509, L514, Y515 were crucial for binding pattern of RdRP with Tetrandrine, Biflavone and Fangchinoline ( Figure 3 and Table 2). Remarkably, binding patterns of spike protein with Isolophirachalcone (R355, Y396, P426, D428, F429, T430, K462, P463, F464, F515), Fangchinoline (R355, Y396, P426, N428, P463, F464, S514, F515, E516) and Tetrandrine (R355, Y396, P426, N428, P463, F464, S514, F515, E516) were exactly similar as all 9 amino acid residues i.e. R355, Y396, P426, N428, P463, F464, S514, F515, E516 were critical for binding with the protein (Figure 4). However, an additional residue T430 were involved in case of Isolophirachalcone (Table 3). P426, D428, T430, P463, F464 and F515 were also critical for binding pattern of Rifampin and SARS-CoV-2 spike protein.

**3.3.Analysis of drug likeness property of top drug candidates**

The most potent MPP, RdRp and spike protein inhibitors (Azobechalcone, Rifampin, Isolophirachalcone, Tetrandrine and Fangchinoline) were investigated in the spheres of physicochemical parameters, lipophilicity, pharmacokinetics and water solubility (Table 5). Lipophilicity, partition coefficient between n-octanol and water (log Po/w) were also calculated by using five widely available predictive models (XLOGP3, WLOGP, MLOGP, SILICOS-IT, iLOGP). GI absorption was lower for all the drug candidates. Azobechalcone, Rifampin, Isolophirachalcone, Tetrandrine and Fangchinoline molecules had no repressive action with the P450 (CYP) isoforms (CYP1A2, CYP2C19, CYP2C9, CYP2D6, CYP3A4). Moreover, no permeant BBB exists among the protein inhibitors of MPP, RdRp and Spike. The compounds sowed water solubility from moderate to high level i.e. 1.30e-16 mg/ml, 1.50e-02 mg/ml, 1.18e-10 mg/ml, 9.78e 09 mg/ml, 4.61e-08 mg/ml, respectively (Table 5). The toxicity analysis of these inhibitors showed that there were no carcinogenic effect and organ toxicity. However, Rifampin, Tetrandrine and Fangchinoline were slightly positive in terms of mutagenesis though Azobechalcone and Isolophirachalcone inhibitors were completely negative. Among the top five inhibitors Azobechalcone was listed in the acute oral toxicity category 2 and rest of them were listed in the category 3.

## 4. Discussion

Global pandemic caused by SARS-CoV-2 has become a major concern due to its excessive infection rate and lethality [51,52,53,54]. Despite huge research regarding the pathogen, no drugs or vaccine has proven satisfactory to combat infections caused by SARS-CoV-2 [55,56]. Several investigational drugs exist, however none of these could treat the patients unquestionably. Moreover, lack of rapid detection procedures made SARS-CoV-2 diagnosis troublesome [57]. Computational approach and drug repurposing strategies hold promise to face such challenges caused by SARS-CoV-2. Hence, in the present study, attempts were taken to suggest probable drug candidates by checking the efficacy of natural inhibitors to inhibit the key proteins of SARS-CoV-2.

The race against the COVID-19 pandemic has allowed the drug repurposing through virtual for finding drugs that could be used for the treatment of COVID-19. Recent studies prioritized MPP inhibitors of SARS-CoV-2 i.e. alpha-ketoamide, Hydroxy, Remdesivir, Chloroquine and Favipiravir to evaluate their potency as drug [58,59,60]. Several in silico approach was also adopted to screen putative drug candidates against SARS-CoV-2 [61,62]. However, all these experiments used either main protease proteins or RNA-dependent RNA polymerase of SARS-CoV-2 as probable drug targets. In this study, we screened potential natural therapeutics against SARS-CoV-2 MPP, RdRp and spike protein by molecular docking approach. Here, Rifampin, Azobechalcone and Azobechalcone were determined as top most drug candidates as they interacted with SARS-CoV-2 MPP, RdRp and spike protein with lowest negative binding energy had the highest negative binding energy (-16.3 kcal/mol, -15.9 kcal /mol and -14.4 kcal/mol respectively. However, comparative analysis revealed the superiority of 5 drug candidates i.e. Azobechalcone, Rifampin, Isolophirachalcone, Tetrandrine, Fangchinoline against SARS-CoV-2 (Table 4). The common drug surface hotspots were analyzed along with modeling of pharmacophore which is very important step for drug discovery. Three amino acid residues i.e. H41, C145, and M165 played the crucial role for the interaction of MPP with its inhibitors (i.e. Rifampin, Amentoflavone, Cepharanthine) (Table 1). Azobechalcone, the top scorer among RdRp inhibitors, Again, the position of F368, L371, L372, A375, W509, L514, Y515 were vital for binding of RdRP with Tetrandrine, Biflavone and Fangchinoline (Table 2). Most importantly, the binding patterns of spike protein with Isolophirachalcone, Fangchinoline and Tetrandrine were significantly similar. This study revealed the possibility of these amino acids to efficiently interact with drugs, though requires validation in wet lab trials. ADME analysis of top drug candidates reveled no undesirable consequences by these compounds. Various physico-chemical parameters, lipophilicity, pharmacokinetics properties and water solubility were determined (Table 5). In addition, no BBB permeant were identified among the top most inhibitors of MPP, RdRp and spike protein. The consequence of association of the natural inhibitory drugs of three key proteins with the cytochrome P450 (CYP) suggests that no substantial inhibition can occur. However, toxicity analysis revealed that Rifampin, Tetrandrine and Fangchinoline can be slightly mutagenic, though there was no possibility for organ toxicity.

The results suggest that Azobechalcone, Rifampin, Isolophirachalcone and Tetrandrine, Fangchinoline could be an option to treat SARS-CoV-2 infections. However, the study employed various computational approaches to screen the potent natural therapeutics and does not involve in-vivo assay. Currently investigational drugs of SARS-CoV-2 are under immense experimental evaluation. Therefore, we suggest clinical trials for the experimental validation of our findings.

## Conclusion

COVID-19 pandemic situation is going to be a worst condition throughout the world. Rapid detection and social distancing are being encouraged at this stage, but we need to search for immediate therapeutic options and effective vaccine candidates for battling this serious health crisis. Drug repurposing approaches could screened out the already approved drugs for reusing against any serious causative agents that are causing health complications. Plant metabolites based repurposed drug molecules could be a promising options against SARS-CoV-2. In the present study, five plantr based therapeutics such as Azobechalcone, Rifampin, Isolophirachalcone, Tetrandrine and Fangchinoline were suggested for potential inhibitors for the Main Protease protein, RNA dependent RNA polymerase and Spike protein of SARS-CoV-2 by using molecular docking based virtual screening study. The study initiated the window towards the thinking of plant based therapy against COVID-19, though extensive research and wet lab validation needs to make it usable for patient.

**Acknowledgements**

Authors would like to acknowledge the Faculty of Biotechnology and Genetic Engineering, Sylhet Agricultural University for the technical support of the project.

**Funding information**

This research did not receive any specific grant from funding agencies in the public, commercial, or not-for-profit sectors.

**Conflict of interest**

Authors declare that they have no conflict of interests.

**Table 1:** Top screened plant metabolites against Main Protease Protein of SARS-CoV-2

| ***No*** | ***Pub Chem ID*** | ***Name*** | ***Binding Energy (Kcal/mol)*** | ***Involved Amino Acids*** |
|---|---|---|---|---|
| 1 | 135398735 | Rifampin | -16.3 | H41, N142, S144, C145, H163,M165, E166, D187, R188, Q189 |
| 2 | 16148290 | Azobechalcone | -14.6 | R131, K137, T196, D197, T198, T199, K236, Y237, N238, L272, G275, M276, L286, L287 |
| 3 | 101630349 | Isolophirachalcone | -13 | K137, D197, T199, Y237, N238, L271, L272, G275, L286, L287, E288, D289, E290 |
| 4 | 5281600 | Amentoflavone | -12.8 | H41, C145, H164, M165, E166, D187, R188, Q189 |
| 5 | 10206 | Cepharanthine | -12.7 | H41, N142, C145, M165, E166, D187, R188, Q189 |

**Table 2:** Top screened plant metabolites against RdRp of SARS-CoV-2

| ***No*** | ***Pub Chem ID*** | ***Name*** | ***Binding Energy (kcal/mol)*** | ***Involved amino acids*** |
|---|---|---|---|---|
| 1 | 16148290 | Azobechalcone | -15.9 | Y32, K47, L49, Y129, H133, N138, D140, T141, S709, T710, D711, K714, K780, N781 |
| 2 | 135398735 | Rifampin | -15.6 | L270, L271, Y273, T324, S325, F326, P328, A282, A283, F396, V398 |
| 3 | 73078 | Tetrandrine | -13.9 | F368, L371, L372, A375, W509, Y515, S518 |
| 4 | 9980790 | Biflavone | -13.7 | F368, L371, L372, A375, W509, L514, Y515 |
| 5 | 73481 | Fangchinoline | -13.7 | F368, L371, L372, A375, W509, L514, Y515, S518 |

**Table 3:** Top screened plant metabolites against Spike Protein of SARS-CoV-2

| ***No*** | ***Pub Chem ID*** | ***Name*** | ***Binding Energy (kcal/mol)*** | ***Involved Amino acid*** |
|---|---|---|---|---|
| 1 | 16148290 | Azobechalcone | -14.4 | L335, C336, F338, G339, F342, N343, V362, A363, D364, V367,L368, S371, F374, W436, N437 |
| 2 | 135398735 | Rifampin | -13.7 | P426, D428, T430, P463, F464, F515, E516, L517, L518 |
| 3 | 101630349 | Isolophirachalcone | -12.8 | R355, Y396, P426, D428, F429, T430, K462, P463, F464, F515 |
| 4 | 73481 | Fangchinoline | -12.6 | R355, Y396, P426, N428, P463, F464, S514, F515, E516 |
| 5 | 73078 | Tetrandrine | -12.5 | R355, Y396, P426, N428, P463, F464, S514, F515, E516 |

**Table 4:** Top screened suggested plant metabolites against Main Protease, RdRpl and Spike Protein of SARS-CoV-2 for battling COVID-19 pandemic

| ***No*** | ***Pub Chem ID*** | ***Name*** | ***Main Protease*** | ***RdRpl*** | ***Spike protein*** |
|---|---|---|---|---|---|
| 1 | 16148290 | Azobechalcone | -14.6 | -15.9 | -14.4 |
| 2 | 135398735 | Rifampin | -16.3 | -15.6 | -13.7 |
| 3 | 101630349 | Isolophirachalcone | -13 | -13.3 | -12.8 |
| 4 | 73078 | Tetrandrine | -12.6 | -13.9 | -12.5 |
| 5 | 73481 | Fangchinoline | -12.3 | -13.7 | -12.6 |

**Table 5:** ADME analysis of top screened suggested plant metabolites against Main Protease, RdRpl and Spike Protein of SARS-CoV-2

| ***Parameter*** | ***Name*** | ***Azobechalcone*** | ***Rifampin*** | ***Isolophirachalcone*** | ***Tetrandrine*** | ***Fangchinoline*** |
|---|---|---|---|---|---|---|
| | ***PubChem ID*** | ***16148290*** | ***135398735*** | ***101630349*** | ***73078*** | ***73481*** |
| Physicochemical parameters | Formula | $C_{90}H_{70}O_{22}$ | $C_{43}H_{58}N_4O_{12}$ | $C_{60}H_{48}O_{15}$ | $C_{38}H_{42}N_2O_6$ | $C_{37}H_{40}N_2O_6$ |
| | Molecular weight | 1503.51 g/mol | 822.94 g/mol | 1009.01 g/mol | 622.75 g/mol | 608.72 g/mol |
| | Molar Refractivity | 415.5 | 234.22 | 277.97 | 186.07 | 181.6 |
| | TPSA | 402.58 Å² | 220.15 Å² | 275.13 Å² | 61.86 Å² | 72.86 Å² |
| Lipophilicity | Log $P_{o/w}$ (iLOGP) | 3.63 | 4.58 | 2.32 | 5.16 | 5.02 |
| | Log $P_{o/w}$ (XLOGP3) | 15.45 | 5.46 | 10.17 | 6.66 | 6.34 |
| | Log $P_{o/w}$ (WLOGP) | 14.52 | 3 | 9.48 | 5.75 | 5.45 |
| | Log $P_{o/w}$ (MLOGP) | 1.77 | 0.14 | 2.02 | 3.73 | 3.55 |
| | Log $P_{o/w}$ (SILICOS-IT) | 12.03 | 2.07 | 7.79 | 6.06 | 5.5 |
| | Consensus Log $P_{o/w}$ | 9.48 | 3.05 | 6.36 | 5.47 | 5.17 |
| Pharmacokinetics | GI absorption | Low | Low | Low | High | High |
| | BBB permeant | No | No | No | No | No |
| | P-gp substrate | Yes | Yes | Yes | No | No |
| | CYP1A2 inhibitor | No | No | No | No | No |
| | CYP2C19 inhibitor | No | No | No | No | No |
| | CYP2C9 inhibitor | No | No | No | No | No |
| | CYP2D6 inhibitor | No | No | No | No | No |
| | CYP3A4 inhibitor | No | No | No | No | No |
| | Log $K_p$ (skin permeation) | -4.50 cm/s | -7.44 cm/s | -5.23 cm/s | -5.37 cm/s | -5.51 cm/s |
| Water Solubility | Log $S$ (SILICOS-IT) | -19.06 | -4.74 | -12.93 | -10.8 | -10.12 |
| | Solubility | 1.30e-16 mg/ml ; 8.67e-20 mol/l | 1.50e-02 mg/ml ; 1.83e-05 mol/l | 1.18e-10 mg/ml ; 1.17e-13 mol/l | 9.78e-09 mg/ml ; 1.57e-11 mol/l | 4.61e-08 mg/ml ; 7.57e-11 mol/l |

**Figure 1:** Top screened plant metabolites for targeting the MPP, RdRp and S protein of SARS-CoV-2

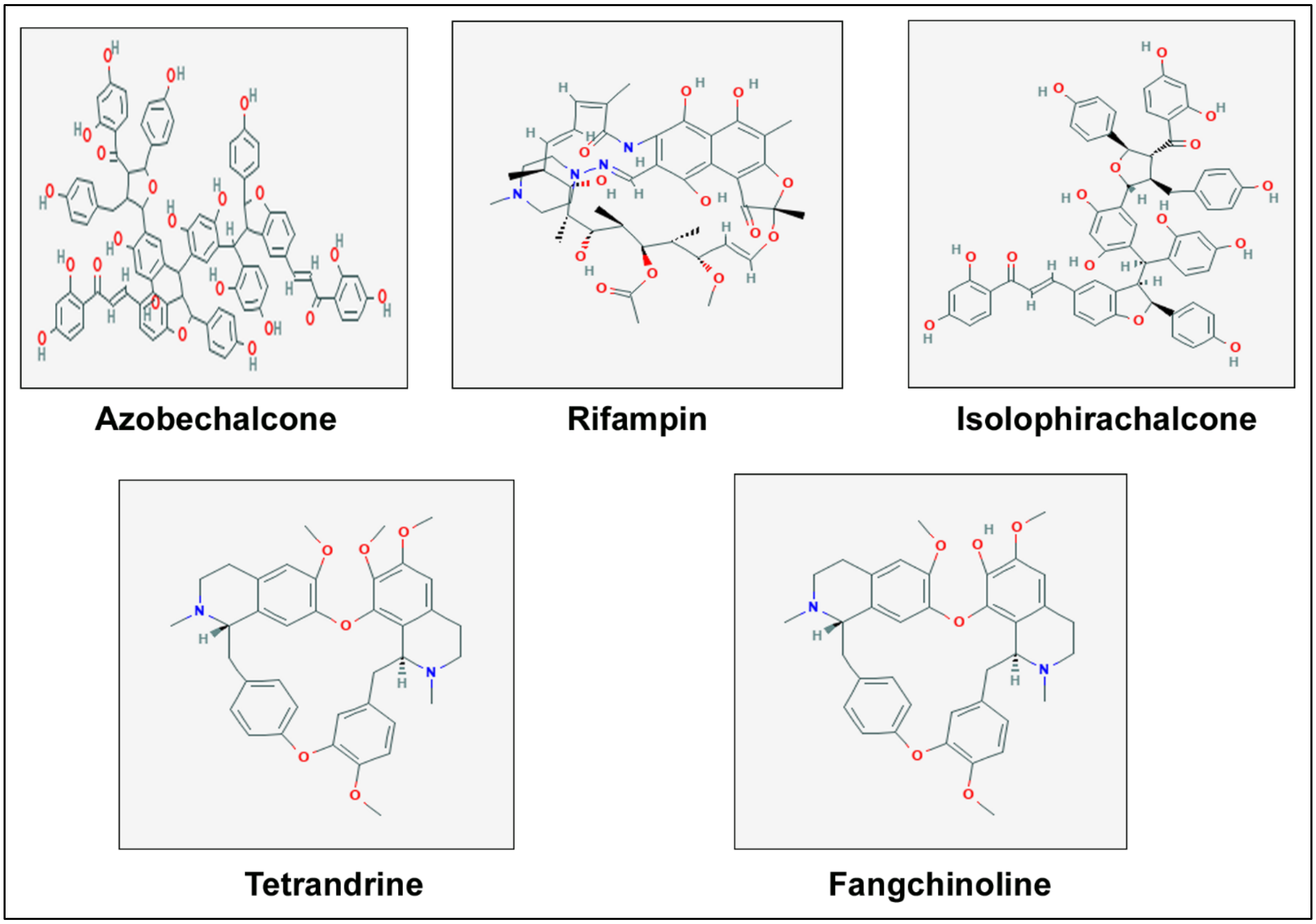

**Figure 2:** Molecular interaction of Rifampin with MPP of SARS-CoV-2 by molecular docking (Binding Energy -16.3 kcal/mol)

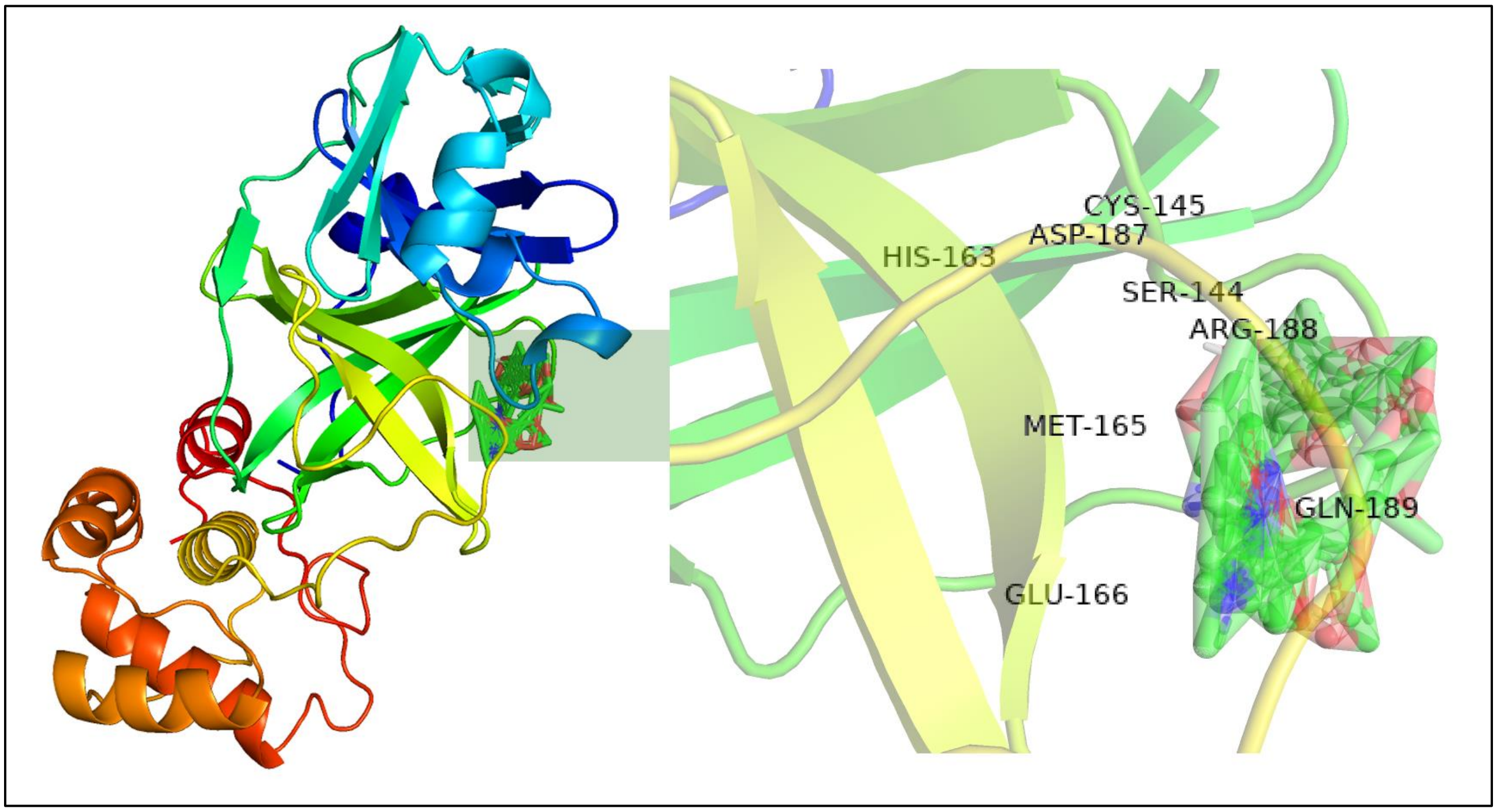

**Figure 3:** Molecular interaction of Azobechalcone with RdRp of SARS-CoV-2 by molecular docking (Binding Energy -15.9 kcal/mol)

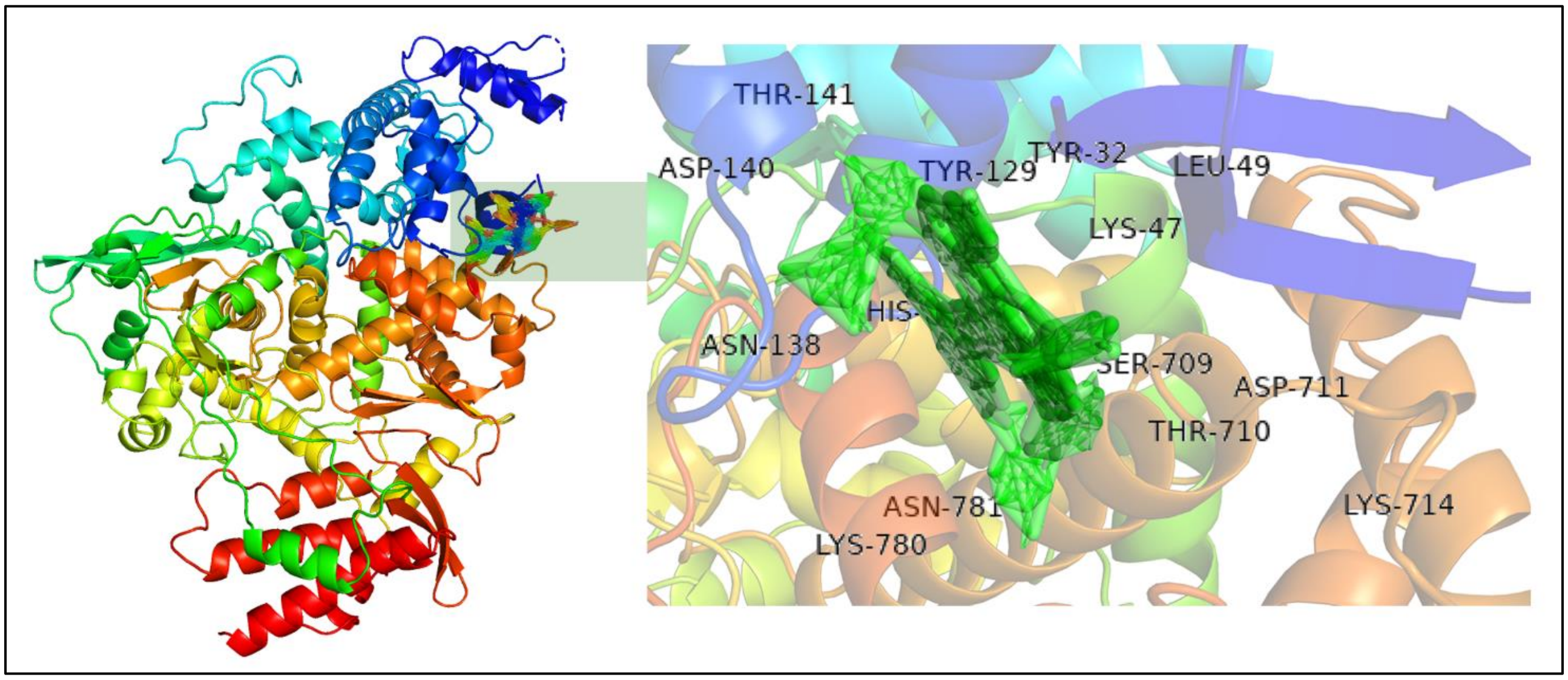

**Figure 4:** Molecular interaction of Azobechalcone with S protein of SARS-CoV-2 by molecular docking (Binding Energy -14.4 kcal/mol)

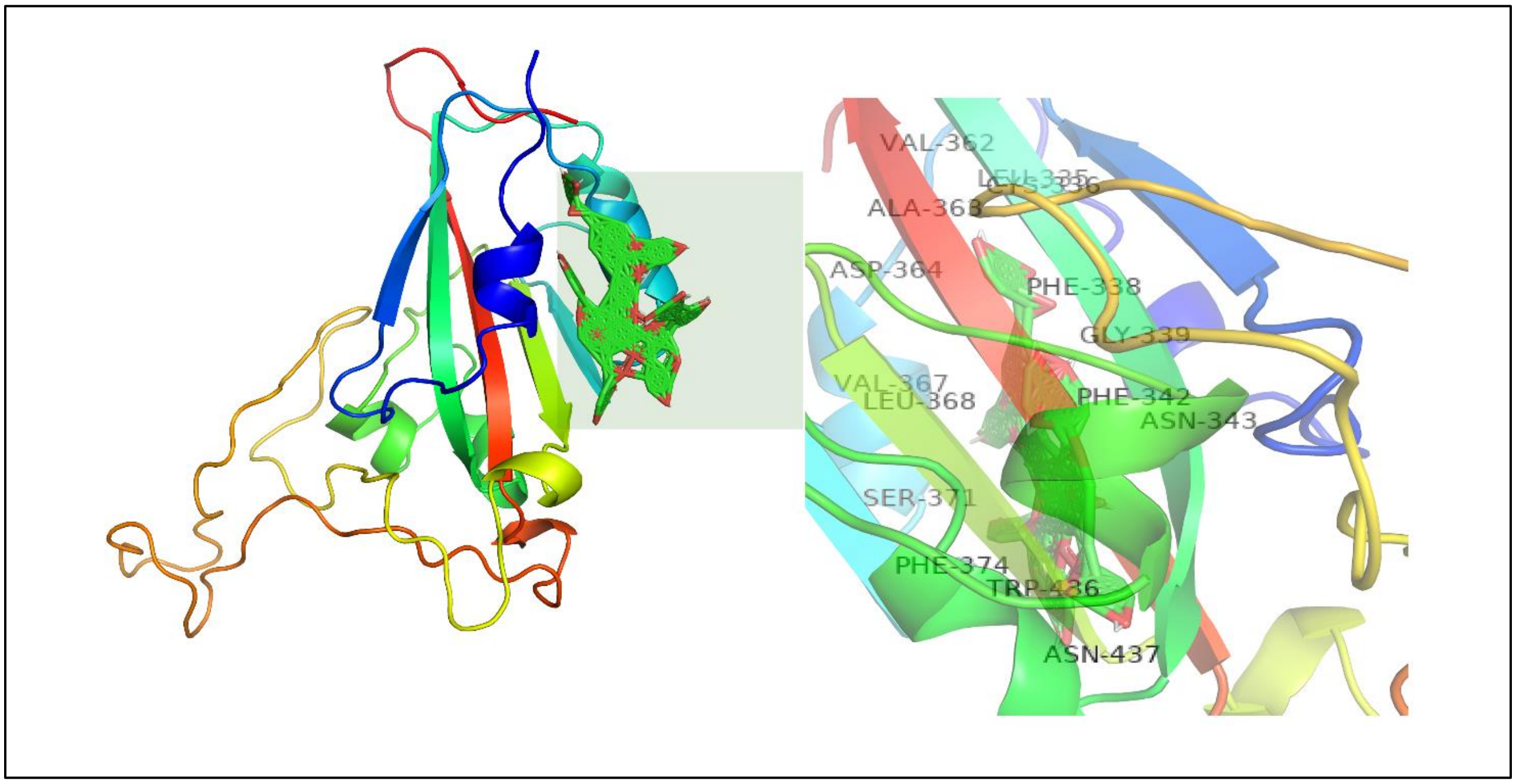

**Supplementary Table 1:** Plant metabolites drug compounds retrieved from PubChem

| ***No*** | ***Name*** | ***Pubchem ID*** | ***Molecular formula*** |
|---|---|---|---|
| 1. | Betulinic Acid | 64971 | $C_{30}H_{48}O_3$ |
| 2. | Hinokinin | 442879 | $C_{20}H_{18}O_6$ |
| 3. | Savinin | 5281867 | $C_{20}H_{16}O_6$ |
| 4. | Curcumin | 969516 | $IC_{21}H_{20}O_6$ |
| 5. | Niclosamide | 4477 | $C_{13}H_8Cl_2N_2O_4$ |
| 6. | Atropine | 174174 | $C_{17}H_{23}NO_3$ |
| 7. | Biopterin | 135449517 | $C_9H_{11}N_5O_3$ |
| 8. | Buchapine | 461150 | $C_{19}H_{23}NO_2$ |
| 9. | Camptothecin | 24360 | $C_{20}H_{16}N_2O_4$ |
| 10. | Canavanin | 439202 | $C_5H_{12}N_4O_3$ |
| 11. | Caffeine | 2519 | $C_8H_{10}N_4O_2$ |
| 12. | Caribine | 441590 | $C_{19}H_{22}N_2O_3$ |
| 13. | Carinatine | 441591 | $C_{17}H_{21}NO_4$ |
| 14. | Chelidonine | 197810 | $C_{20}H_{19}NO_5$ |
| 15. | Cordycepin | 6303 | $C_{10}H_{13}N_5O_3$ |
| 16. | Cryptopleurine | 92765 | $C_{24}H_{27}NO_3$ |
| 17. | O-Demethyl-Buchenavianine | 5352072 | $C_{21}H_{21}NO_4$ |
| 18. | Emetine | 10219 | $C_{29}H_{40}N_2O_4$ |
| 19. | Fagaronine | 40305 | $C_{21}H_{20}NO_4^+$ |
| 20. | Harmaline | 3564 | $C_{13}H_{14}N_2O$ |
| 21. | Harmine | 5280953 | $C_{13}H_{12}N_2O$ |
| 22. | Hypoxanthine | 135398638 | $C_5H_4N_4O$ |
| 23. | Lycorine | 72378 | $C_{16}H_{17}NO_4$ |
| 24. | Michellamines D | 403970 | |
| 25. | Michellamines F | 403969 | |
| 26. | 10-Methoxycamptothecin | 97283 | $C_{21}H_{18}N_2O_5$ |
| 27. | Odorinol | 6440456 | $C_{18}H_{24}N_2O_3$ |
| 28. | Oliverine | 24198103 | $C_{20}H_{22}ClNO_4$ |
| 29. | Oxostephanine | 343547 | $C_{18}H_{11}NO_4$ |
| 30. | Pachystaudine | 513596 | $C_{19}H_{19}NO_4$ |
| 31. | Papaverine | 4680 | $C_{20}H_{21}NO_4$ |
| 32. | Psychotrine | 65380 | $C_{28}H_{36}N_2O_4$ |
| 33. | Rifampin | 135398735 | $C_{43}H_{58}N_4O_{12}$ |
| 34. | Schumannificine | 184890 | |
| 35. | Solasonine | 119247 | $C_{45}H_{73}NO_{16}$ |
| 36. | Taspine | 215159 | $C_{20}H_{19}NO_6$ |
| 37. | Homonojirimycin | 159496 | $C_7H_{15}NO_5$ |
| 38. | Aranotin | 10412012 | $C_{20}H_{18}N_2O_7S_2$ |
| 39. | Gliotoxin | 6223 | $C_{13}H_{14}N_2O_4S_2$ |
| 40. | Ochropamine | 134716677 | $C_{22}H_{26}N_2O_3$ |

| | Name | Pubchem ID | Molecular formula |
|---|---|---|---|
| 41. | Epi-16-Ochropamine | 6445875 | $C_{22}H_{27}ClN_2O_3$ |
| 42. | Glaucine Fumarate | 16754 | $C_{21}H_{25}NO_4$ |
| 43. | N-Methyllaurotetanine | 16573 | $C_{20}H_{23}NO_4$ |
| 44. | Isoboldine | 133323 | $C_{19}H_{21}NO_4$ |
| 45. | Nuciferine Hcl | 53326062 | $C_{19}H_{22}ClNO_2$ |
| 46. | Indigo | 10215 | $C_{16}H_{10}N_2O_2$ |
| 47. | Sinigrin | 23682211 | $C_{10}H_{16}KNO_9S_2$ |
| 48. | Aloeemodin | 10207 | $C_{15}H_{10}O_5$ |
| 49. | Hesperetin | 72281 | $C_{16}H_{14}O_6$ |
| 50. | Hirsutenone | 637394 | $C_{19}H_{20}O_5$ |
| 51. | Platyphyllenone | 23786382 | $C_{19}H_{20}O_3$ |
| 52. | Platyphyllone | 13347313 | $C_{19}H_{22}O_4$ |
| 53. | Platyphyllonol-5-Xylopyranoside | 46230810 | $C_{24}H_{30}O_8$ |
| 54. | Hirsutanonol | 9928190 | $C_{19}H_{22}O_6$ |
| 55. | Oregonin | 14707658 | $C_{24}H_{30}O_{10}$ |
| 56. | Rubranol | 10088141 | $C_{19}H_{24}O_5$ |
| 57. | Rubranoside B | 24011643 | $C_{24}H_{32}O_9$ |
| 58. | Rubranoside A | 10097263 | $C_{25}H_{34}O_{10}$ |
| 59. | Biflavone | 9980790 | $C_{30}H_{18}O_4$ |
| 60. | Amentoflavone | 5281600 | $C_{30}H_{18}O_{10}$ |
| 61. | Apigenin | 5280443 | $C_{15}H_{10}O_5$ |
| 62. | Luteolin | 5280445 | $C_{15}H_{10}O_6$ |
| 63. | Quercetin | 5280343 | $C_{15}H_{10}O_7$ |
| 64. | Tomentin A | 71659627 | $C_{25}H_{30}O_7$ |
| 65. | Tomentin B | 71659628 | $C_{26}H_{32}O_7$ |
| 66. | Tomentin C | 71659765 | $C_{27}H_{34}O_8$ |
| 67. | Tomentin D | 71659766 | $C_{27}H_{34}O_8$ |
| 68. | Tomentin E | 71659767 | $C_{26}H_{32}O_8$ |
| 69. | Chalcones | D047188 | |
| 70. | Coumarins | 54678486 | $C_{19}H_{16}O_4$ |
| 71. | Herbacetin | 5280544 | $C_{15}H_{10}O_7$ |
| 72. | Rhoifolin | 5282150 | $C_{27}H_{30}O_{14}$ |
| 73. | Pectolinarin | 168849 | $C_{29}H_{34}O_{15}$ |
| 74. | Lycorine | 72378 | $C_{16}H_{17}NO_4$ |
| 75. | Tryptanthrin | 73549 | $C_{15}H_8N_2O_2$ |
| 76. | Scutellarein | 5281697 | $C_{15}H_{10}O_6$ |
| 77. | Myricetin | 5281672 | $C_{15}H_{10}O_8$ |
| 78. | Glycyrrhizin | 14982 | $C_{42}H_{62}O_{16}$ |

| No | Name | Pubchem ID | Molecular formula |
|---|---|---|---|
| 79. | Herbacetin | 5280544 | $C_{15}H_{10}O_7$ |
| 80. | Quercetin | 5280343 | $C_{15}H_{10}O_7$ |
| 81. | Helichrysetin | 6253344 | $C_{16}H_{14}O_5$ |
| 82. | Tetrandrine | 73078 | $C_{38}H_{42}N_2O_6$ |
| 83. | Fangchinoline | 73481 | $C_{37}H_{40}N_2O_6$ |
| 84. | Cepharanthine | 10206 | $C_{37}H_{38}N_2O_6$ |
| 85. | Rhoifolin | 5282150 | $C_{27}H_{30}O_{14}$ |
| 86. | Pectolinarin | 168849 | $C_{29}H_{34}O_{15}$ |
| 87. | Epigallocatechin | 65064 | $C_{22}H_{18}O_{11}$ |
| 88. | Gallocatechin Gallate | 199472 | $C_{22}H_{18}O_{11}$ |
| 89. | Emodin | 3220 | $C_{15}H_{10}O_5$ |
| 90. | Baicalin | 64982 | $C_{21}H_{18}O_{11}$ |
| 91. | Saikosaponins | 107793 | $C_{42}H_{68}O_{13}$ |
| 92. | Tetra-O-Galloyl-B-D-Glucose | 471531 | $C_{34}H_{28}O_{22}$ |
| 93. | Kaempferol | 5280863 | $C_{15}H_{10}O_6$ |
| 94. | Juglanin | 5318717 | $C_{20}H_{18}O_{10}$ |
| 95. | Tiliroside | 5320686 | $C_{30}H_{26}O_{13}$ |
| 96. | Afzelin | 5316673 | $C_{21}H_{20}O_{10}$ |
| 97. | Naringenin | 932 | $C_{15}H_{12}O_5$ |
| 98. | Genistein | 5280961 | $C_{15}H_{10}O_5$ |
| 99. | Castanospermine | 54445 | $C_8H_{15}NO_4$ |
| 100. | Australine | 442628 | $C_8H_{15}NO_4$ |
| 101. | Periformyline | 11969538 | $C_{21}H_{22}N_2O_4$ |
| 102. | Perivine | 6473766 | $C_{20}H_{22}N_2O_3$ |
| 103. | Vincaleucoblastine | 241902 | $C_{46}H_{60}N_4O_{13}S$ |
| 104. | Columbamine | 72310 | $C_{20}H_{20}NO_4^+$ |
| 105. | Berberine | 2353 | $C_{20}H_{18}NO_4^+$ |
| 106. | Palmitine | 19009 | $C_{21}H_{22}NO_4^+$ |
| 107. | Narciclasine | 72376 | $C_{14}H_{13}NO_7$ |
| 108. | Lycoricidine | 73065 | $C_{14}H_{13}NO_6$ |
| 109. | Pancratistatin | 441597 | $C_{14}H_{15}NO_8$ |
| 110. | 7-deoxypancratistatin | 443741 | $C_{14}H_{15}NO_7$ |
| 111. | Isonarciclasine | 5479394 | $C_{14}H_{13}NO_7$ |
| 112. | cis-Dihydronarciclasine | 3000372 | $C_{14}H_{15}NO_7$ |
| 113. | Pretazettine | 73360 | $C_{18}H_{21}NO_5$ |
| 114. | Buxamine E | 442970 | $C_{26}H_{44}N_2$ |
| 115. | Cyclobuxamine H | 101281345 | $C_{24}H_{42}N_2O$ |
| 116. | 5-hydroxynoracronycine | 5378702 | $C_{19}H_{17}NO_4$ |
| 117. | Acrimarine F | 5491701 | $C_{31}H_{29}NO_8$ |

**Supplementary Table 2:** Docking results of all plant metabolites drug compounds against three key proteins of SARS-CoV-2

| *Pub Chem ID* | *Binding energy against Main Protease (kcal/mol)* | *Binding energy against RdRp (kcal/mol)* | *Binding energy against Spike protein (kcal/mol)* |
|---|---|---|---|
| CID 16148290 | -14.6 | -15.9 | -14.4 |
| CID 135398735 | -16.3 | -15.6 | -13.7 |
| CID 101630349 | -13 | -13.3 | -12.8 |
| CID 73481 | -12.3 | -13.7 | -12.6 |
| CID 73078 | -12.6 | -13.9 | -12.5 |
| CID 10206 | -12.7 | -13.5 | -12.3 |
| CID 64971 | -12.3 | -13.1 | -11.9 |
| CID 5281600 | -12.8 | -13.3 | -11.6 |
| CID 9980790 | -12.4 | -13.7 | -11.5 |
| CID 14982 | -12.6 | -13.6 | -11.4 |
| CID 107793 | -11.2 | -12.4 | -11.3 |
| CID 119247 | -10.9 | -12 | -11.1 |
| CID 5281694 | -11.6 | -11.7 | -11 |
| CID 5281599 | -12.1 | -12.1 | -10.9 |
| CID 97283 | -10.6 | -10.2 | -10.8 |
| CID 24360 | -10.5 | -10.4 | -10.6 |
| CID 5480834 | -12.1 | -11.5 | -10.5 |
| CID 5491701 | -11.9 | -11.8 | -10.4 |
| CID 72412 | -11.6 | -12.1 | -10.2 |
| CID 455249 | -10.8 | -12.1 | -10.2 |
| CID 10412012 | -11 | -10.6 | -10.1 |
| CID 102056152 | -10.5 | -11.5 | -9.9 |
| CID 54675769 | -10.4 | -10.6 | -9.9 |
| CID 5464386 | -11.5 | -11.6 | -9.8 |
| CID 6473766 | -10 | -10.3 | -9.8 |
| CID 168849 | -10.2 | -10.7 | -9.7 |
| CID 168849 | -10.6 | -10.1 | -9.7 |
| CID 11969538 | -9.4 | -9.9 | -9.7 |
| CID 71659627 | -9 | -9.6 | -9.7 |
| CID 442970 | -9.8 | -10.7 | -9.5 |
| CID 5352072 | -10.5 | -10 | -9.5 |
| CID 64982 | -9.8 | -9.9 | -9.5 |
| CID 71659765 | -9.1 | -10.2 | -9.3 |
| CID 54678486 | -9.4 | -9.8 | -9.3 |
| CID 5281867 | -8.6 | -9.8 | -9.3 |
| CID 442879 | -8.9 | -8.8 | -9.3 |
| CID 101281345 | -9.5 | -10.7 | -9.2 |
| CID 461150 | -8.9 | -9.7 | -9.2 |
| CID 441590 | -9.5 | -9.4 | -9.1 |
| CID 5282150 | -9.9 | -11.2 | -9 |
| CID 442428 | -9.9 | -11.2 | -9 |
| CID 5282150 | -10.1 | -11.1 | -9 |
| CID 134716677 | -9.5 | -10.4 | -9 |
| CID 71659628 | -9 | -9.7 | -9 |
| CID 5316673 | -10.3 | -9.7 | -8.9 |
| CID 5318717 | -9.2 | -9.7 | -8.9 |
| CID 71659766 | -8.2 | -9.1 | -8.9 |
| CID 5378702 | -9 | -9.8 | -8.8 |
| CID 5280805 | -11.5 | -10.6 | -8.7 |
| CID 5280804 | -9.6 | -9.9 | -8.7 |
| CID 5320686 | -8.5 | -9.9 | -8.6 |
| CID 122851 | -10 | -9.7 | -8.6 |
| CID5280637 | -9.7 | -9.4 | -8.6 |
| CID 199472 | -9.2 | -9.5 | -8.5 |
| CID 65064 | -9.2 | -9.4 | -8.5 |
| CID 6440456 | -8.4 | -9.4 | -8.5 |
| CID 10219 | -8.8 | -9.1 | -8.5 |
| CID 71659767 | -8.6 | -9 | -8.5 |
| CID 14707658 | -10.2 | -9.7 | -8.4 |
| CID 442882 | -9.7 | -9.7 | -8.4 |
| CID 513596 | -8.7 | -9.2 | -8.4 |
| CID 40305 | -8.2 | -9.1 | -8.4 |
| CID 6445875 | -8.3 | -8.6 | -8.4 |
| CID 73360 | -8.6 | -9 | -8.3 |
| CID 72378 | -8.6 | -8.4 | -8.3 |
| CID 471531 | -11.1 | -11 | -8.2 |
| CID 184890 | -9 | -9.9 | -8.2 |
| CID 65380 | -9.2 | -8.8 | -8.2 |
| CID 5281697 | -8.5 | -8.7 | -8.2 |
| CID 5281672 | -8.2 | -8.6 | -8.2 |
| CID 73065 | -8.2 | -8.6 | -8.2 |
| CID 2353 | -8.5 | -8.5 | -8.2 |
| CID 24198103 | -7.9 | -8.4 | -8.2 |
| CID 5479394 | -8.7 | -9.3 | -8.1 |
| CID 443741 | -8.5 | -9.3 | -8.1 |
| CID 40305 | -8.2 | -9.1 | -8.1 |
| CID 343547 | -9 | -9 | -8.1 |
| CID 439533 | -8.1 | -8.9 | -8.1 |
| CID 10215 | -8.1 | -8.8 | -8.1 |
| CID 72376 | -8.7 | -8.7 | -8.1 |
| CID 5280343 | -8.1 | -8.7 | -8.1 |
| CID 5280343 | -8.1 | -8.7 | -8.1 |
| CID 3000372 | -8.4 | -8.6 | -8.1 |
| CID 5281680 | -8.3 | -8.5 | -8.1 |
| CID 5281614 | -8.1 | -8.3 | -8.1 |
| CID 637394 | -7.4 | -7.8 | -8.1 |
| CID 10088141 | -7.4 | -7.7 | -8.1 |
| CID 92765 | -8.8 | -9.7 | -8 |
| CID 441597 | -8.6 | -9.3 | -8 |
| CID 3220 | -8.6 | -9.1 | -8 |
| CID 73549 | -8.1 | -8.7 | -8 |
| CID 72281 | -8 | -8.5 | -8 |
| CID 6223 | -8.5 | -8.4 | -8 |
| CID 5280445 | -8.2 | -8.3 | -8 |
| CID 932 | -8.2 | -8.6 | -7.9 |
| CID 5280443 | -8.2 | -8.6 | -7.9 |
| CID 23786382 | -7 | -8.6 | -7.9 |
| CID 24011643 | -8.8 | -8.5 | -7.9 |
| CID 5281670 | -8.2 | -8.5 | -7.9 |
| CID 5489090 | -9.2 | -8.3 | -7.9 |
| CID 5280443 | -8.2 | -8.3 | -7.9 |
| CID 9928190 | -7.5 | -8.3 | -7.9 |
| CID 5317756 | -8.5 | -9.6 | -7.8 |
| CID 10207 | -8 | -8.6 | -7.8 |
| CID 5280681 | -8.1 | -8.4 | -7.8 |
| CID 5280544 | -8 | -8.3 | -7.8 |
| CID 5280544 | -8 | -8.3 | -7.8 |
| CID 16573 | -7.8 | -8 | -7.8 |
| CID 969516 | -7.1 | -6.9 | -7.8 |
| CID 5280862 | -8 | -8.1 | -7.7 |
| CID 5281603 | -8.2 | -8 | -7.7 |
| CID 5280961 | -8.1 | -8 | -7.7 |
| CID 5280863 | -7.9 | -8 | -7.7 |
| CID 440832 | -7.6 | -7.9 | -7.7 |
| CID 4477 | -7.9 | -7.4 | -7.7 |
| CID 13347313 | -7.1 | -7.2 | -7.7 |
| CID 441591 | -8 | -8.6 | -7.6 |
| CID 10097263 | -9 | -8.4 | -7.6 |
| CID 133323 | -8.2 | -8.4 | -7.6 |
| CID 53326062 | -6.9 | -8.4 | -7.6 |
| CID 5316900 | -7.9 | -8 | -7.6 |
| CID 19009 | -8.1 | -7.9 | -7.6 |
| CID 5281677 | -8.1 | -7.8 | -7.6 |
| CID 5281677 | -8.1 | -7.8 | -7.6 |
| CID 5280378 | -7.6 | -7.8 | -7.6 |
| CID 44259819 | -9 | -8.8 | -7.5 |
| CID 46230810 | -6.8 | -8.7 | -7.4 |
| CID 197810 | -8 | -8.2 | -7.4 |
| CID 189065 | -8.1 | -8.1 | -7.4 |
| CID 72310 | -7.9 | -8 | -7.3 |
| CID 72310 | -7.9 | -8 | -7.3 |
| CID 5459184 | -7.6 | -7.8 | -7.3 |
| CID 5281704 | -7.8 | -7.6 | -7.3 |
| CID 16754 | -7.9 | -8.1 | -7.2 |
| CID 6253344 | -6.8 | -7.4 | -7.2 |
| CID 23682211 | -7.5 | -7.7 | -7.1 |
| CID 215159 | -7.8 | -8 | -7 |
| CID 4680 | -7 | -7.1 | -7 |
| CID 5280953 | -6.4 | -6.7 | -6.8 |
| CID 119586 | -6.9 | -7.4 | -6.7 |
| CID 174174 | -7 | -6.8 | -6.7 |
| CID 3564 | -6.6 | -7.5 | -6.6 |
| CID 6303 | -6.7 | -7.1 | -6.6 |
| CID 135449517 | -6.5 | -6.6 | -6.6 |
| CID 5281541 | -6.2 | -6.7 | -6.1 |
| CID 54445 | -6.6 | -7.1 | -6 |
| CID 159496 | -5.6 | -7.1 | -5.7 |
| CID 442628 | -6 | -6.4 | -5.6 |
| CID 2519 | -5.7 | -6.1 | -5.4 |
| CID 403969 | -5.2 | -5.8 | -5.2 |
| CID 135398638 | -5.7 | -6 | -4.9 |
| CID 439202 | -5.1 | -5.3 | -4.9 |
| CID 241902 | -3.8 | -4.5 | -3.4 |